\documentclass[%
 reprint,
 amsmath,amssymb,
 aps,
 pre
]{revtex4}

\usepackage{graphicx}
\usepackage{dcolumn}
\usepackage{bm}

\begin{document}

\preprint{APS/123-QED}

\title{Wetting Transition in the Two-Dimensional Blume-Capel Model: A Monte Carlo study}
\thanks{kurt.binder@uni-mainz.de, ealbano@iflysib.unlp.edu.ar}%

\author{Ezequiel V. Albano}
 \affiliation{Instituto de F\'isica de L\'iqidos y Sistemas Biol\'ogicos (IFLYSIB).
 CCT-CONICET La Plata, UNLP.
 Calle 59 Nro. 789, (1900) La Plata, Argentina.} %


\author{Kurt Binder}
\affiliation{Institut f\"ur Physik, Johannes Gutenberg Universit\"at Mainz,\\
Staudinger Weg 7, 55099 Mainz, Germany}

\date{\today}

\begin{abstract}
The wetting transition of the Blume-Capel model is studied by a finite-size scaling analysis
of $L \times M$ lattices where competing boundary fields $\pm H_1$ act on the first
row or last row of the $L$ rows in the strip, respectively. We show that using the appropriate
anisotropic version of finite size scaling, critical wetting in $d=2$ is equivalent to a ``bulk'' critical
phenomenon with exponents $\alpha =-1$, $\beta =0$, and $\gamma=3$. These concepts
are also verified for the Ising model. For the Blume-Capel model it is found that the field
strength $H_{1c} (T)$ where critical wetting occurs goes to zero when the bulk second-order transition
is approached, while $H_{1c}(T)$ stays nonzero in the region where in the bulk a first-order
transition from the ordered phase, with nonzero spontaneous magnetization, to the
disordered phase occurs. Interfaces between coexisting phases then show interfacial enrichment
of a layer of the disordered phase which exhibits in the second order case a finite thickness only.
A tentative discussion of the scaling behavior of the wetting phase diagram near the tricritical 
point also is given.

\pacs{Valid PACS appear here}
PACS: 68.08.Bc, 68.35.Rh, 64.60F-, 05.70.Fh

\end{abstract}
\maketitle


\section{Introduction}
Understanding interfacial phenomena at boundaries \cite{1,2,3,4} such as free surfaces of solids, or
walls confining fluids, etc., is a topic of very high interest in current condensed matter research,
driven by important applications, e.g. in micro- or nanofluidics, design of smart nanomaterials
\cite{5,6,7,8}, etc. But it is also of fundamental interest as a problem of statistical mechanics. It is
only the latter aspect that will be in our focus here, and hence we shall not deal here with wetting
properties \cite{9,10,11,12,13,14,15,16,17} of specific materials, but rather consider wetting in the
framework of a simple lattice model with nearest neighbor interactions and short-range forces due to the
walls.

In this spirit, a large amount of work has been devoted to the study of simple Ising models with
boundary fields; see e.g. \cite{18} for a review, 
and \cite{19,20,21,22,23,24,25,26,27,28,29,30,31,32,33,34,35,36,37,38,39,40} for some pertinent 
original work. Spin reversal symmetry of this generic model
implies that phase coexistence between the ordered phases with positive and negative spontaneous
magnetization occurs at zero bulk field (H=0) in the thermodynamic limit in the bulk; the same holds
still true when one considers a thin film of a finite thickness $L$ with surfaces at which boundary fields
$H_1$, $H_L$ act, provided one considers the antisymmetric situation $H_1=-H_L$, avoiding ``capillary
condensation'' \cite{18,20,22}, i.e.~a shift of bulk coexistence to nonzero value of the bulk field due to
boundary effects. The latter phenomena is interesting in its own right, e.g.~\cite{41}, and can also
be studied in Ising models choosing boundary fields $H_1=H_L$, e.g.~\cite{42,43,44}, but this
problem will not be followed up here.

In principle, the study of a wetting transition, where the boundary induces the formation of a
macroscopically thick film of the phase that it prefers, coexisting with the other phase separated
by an interface at a macroscopic distance from the boundary, requires to consider the limit
$L \rightarrow \infty$. For any finite $L$ in the antisymmetric thin film geometry, or
thin strip geometry, when one considers $d=2$ dimensions rather than $d=3$, one does not find a
wetting transition, but rather the "interface 
localization-delocalization transition'' \cite{18,27,28,29,30,31,32,33,37,38,39}
: for temperatures above $T_c(L, H_1)$ but below the transition temperature $T_{cb}$ of the bulk
Ising model, the interface between the coexisting phases is ``delocalized", freely fluctuating in the
center of the film (or strip, respectively). Here, by center we mean the plane (or line) ($L+1)/2$, 
if we label the layers (rows) of the Ising lattice parallel to the boundaries from $n=1$ to $n=L$. 
However, for $T<T_c(L,H_1)$ the interface is tightly bound, and hence
``localized'', near one of the two boundaries. One predicts, however, that for large $L$
the transition temperature $T_c(L,H_1)$ converges rapidly to the wetting transition
temperature $T_w(H_1)$ of the semi-infinite system \cite{29,30}, and this prediction is
consistent with the numerical simulations \cite{27,28,31,32,33,38,39}.

One aspect of wetting in the Ising model that found great attention is the behavior near
bulk criticality. One can show that for a second-order wetting transition the inverse
function $H_{1c}(T)$ of $T_w(H_1)$ behaves as \cite{21}

\begin{equation} \label{eq1}
H_{1c} (T) \propto (T_{cb} - T) ^{\Delta _1}  \quad  ,
\end{equation}

\noindent where $\Delta_1$ is the critical exponent that controls the scaling behavior with the surface
field $H_1$ near bulk criticality \cite{45,46,47}. While $\Delta_1$ in $d=3$
is only known approximately from numerical work, in $d=2$ Abraham's exact solution
for the Ising model \cite{19} implies $\Delta_1=\frac{1}{2}$. 
Note that on the square lattice with exchange constant $J$
between nearest neighbors the wetting transition occurs for $H_{1c}$ given as the
solution of

\begin{eqnarray} \label{eq2}
&&\exp (2 J/k_BT) [\cosh (2 J/k_BT) - \cosh (2 H_{1c}/k_BT)]\nonumber\\
&&=\sinh (2 J/k_BT) \quad ,
\end{eqnarray}

\noindent and the bulk critical point occurs at $\exp(2J/k_B T_{cb})= \sqrt{2} +1 $ \cite{48}.

An intriguing question is, of course, to study how the wetting behavior gets modified
if the bulk phase transition changes its character. E.g., a simple extension of the
Ising model is the Blume-Capel model \cite{49,50}, where each lattice site (i)
can be in one of three states $S_i=+1$, $S_i=0$, $S_i=-1$ and the
Hamiltonian becomes

\begin{equation} \label{eq3}
\mathcal{H}=-J \sum\limits_{\langle i,j\rangle} S_iS_j + D \sum\limits_i S_i^2 \quad  ,
\end{equation}

\noindent  where the ``crystal field'' $D$ controls the density of ``vacancies'', 
i.e. sites with $S_i=0$.

When $D \rightarrow -\infty$ vacancies are excluded, and the model reduces to
the standard two state Ising model, but when $D/J$ is of order unity,
one has a nontrivial bulk phase diagram with a tricritical point \cite{51}
at $D=D_t$. For the square lattice, this tricritical point occurs
at $D_t/J \tilde{=} 1.965(5)$ and $k_BT_t/J \tilde{=} 0.609$ (5).
For $D >  D_t$ the transition at $T_c(D)$ is first order: two phases
with opposite nonzero magnetization and the disordered phase (with
zero magnetization) coexist.

From the universality principle, we expect that  Eq.~(\ref{eq1}) with 
$\Delta_1=\frac{1}{2}$ would also hold in the second-order region of the 
Blume-Capel model for $D < D_t$, but to our knowledge this assumption 
has not yet been tested. Providing evidence for this assumption is
one of the goals of the present paper. Right at the tricritical point,
however, the behavior should be different: instead of Eq.~(\ref{eq1}) 
we now expect 

\begin{equation} \label{eq4}
H_{1c} (D_t, T) \propto (T_t - T) ^{\Delta_{1t}} \quad,
\end{equation}

\noindent where the exponent $\Delta_{1t}$ in $d=3$ is expected to be $\Delta_{1t}=1/4$
\cite{52}, the prediction due to Landau mean-field theory, which is expected
to hold in $d=3$, apart from logarithmic corrections \cite{51}. While in
$d=2$ the tricritical exponents in the bulk are not those of mean-field theory
\cite{51}, for the Ising model universality class to which the
tricritial point of the Blume-Capel model belongs \cite{51}, they are exactly
known from conformal invariance \cite{53,54,55}. E.g. in standard notation
of critical exponents \cite{56,57}) $\alpha_t=8/9$, 
$\beta_t=1/24$, $\gamma_t=37/36$, $\nu_t=5/9$, etc.
To our knowledge, $\Delta_{1t}$ for this problem is not yet known, however.

When one considers the wetting transition for the part of the transition
line $T_{cb}(D)$ that is first order $(D > D_t)$, the wetting transition
$H_{1c}(T)$ ends at this line at a nonzero value of the surface field,

\begin{equation} \label{eq5}
H^*_{1c} (D) =H_{1c} (D, T=T_{cb}(D)) \quad .
\end{equation}

We expect that another power law will exist when $D$ approaches $D_t$ from below,

\begin{equation} \label{eq6}
H^*_{1c} (D) \propto (D - D_t)^\zeta \quad ,
\end{equation}

\noindent but again we are not aware of predictions relating to this critical 
exponent $\zeta$.  The
evaluation of  $\zeta$ from numerical work is a very challenging task, because 
it requires to locate both the tricritical point and the wetting transitions
with very high accuracy. 

 Finally, a very interesting aspect relates to the character of the interface
 between the coexisting phases with positive and negative magnetization, when
 one approaches the bulk phase transition: then interfacial adsorption of the third phase
 (the disordered phase) at the interface can occur \cite{58,59,60}. If we denote
 the coexisting bulk phases as $A$ and $B$, approaching the bulk transition line in the
 second-order region one expects an interfacial wetting transition $A|B \rightarrow A|DO|B$,
 where $DO$ stands symbolically for an intruding layer of the disordered phase
 having predominantly states where $S_i=0$ dominates at the interface. To quantify
 this effect one considers the net adsorption defined by 

 \begin{equation} \label{eq7}
 W_o(T) = N ^{-1} \sum\limits_{i=1}^M [ \langle \delta_{0,S_i} \rangle _{1|-1} -
 \langle \delta_{0,S_i} \rangle_{1|1}]  \quad ,
 \end{equation}

 \noindent where $N$ is the total number of lattice sites, $M$ is the number of lattice sites 
per row parallel to the walls, $\delta_{\alpha, \beta}$ is the Kronecker symbol, 
and $\langle \cdots \rangle_{1|-1}$
 a statistical average in the presence of an interface ($A=1$, $B=-1$), while
 $\langle \cdots \rangle_{1|1}$ is the corresponding average for an equivalent
 system but without interfaces. Since $\delta_{0,S_i} =1-S_i^2$, one can conclude
 also that $W_0(T) = (\partial \sigma_{AB} /\partial D)$ where $\sigma_{AB}$ is the
 interfacial tension for the considered interface. Since for $D < D_t$ the bulk phase
 transition of the Blume-Capel model falls in the Ising universality class, we know
 that $\sigma_{AB} \propto (1-T/T_{cb}(D))^\mu$ with $\mu=(d-1) \nu=1$ for the Blume-Capel
 model. Writing $t=1-T/T_{cb} (D)$, we find

 \begin{equation} \label{eq8}
 W_0(T)=\Big(\frac{\partial \sigma_{AB}}{\partial t}\Big) \Big(\frac{\partial t}{\partial D}\Big) \propto t ^{\mu-1} ={ \rm const} \quad ;
 \end{equation}

 \noindent thus if we define a divergence of interfacial adsorption via

 \begin{equation} \label{eq9}
 W_0(T) \propto t^{- \omega} \quad ,
 \end{equation}

 \noindent we find $\omega =0$ for $D < D_t$, i.e.~in the Ising-like regime there should be
 no critical divergence of the net adsorption. However, right at the tricritical
 point, the behavior should be different: from $\nu=5/9$ as quoted above we conclude
 $W_0(T) \propto t ^{- \omega}$ with $\omega=4/9$ for $D=D_t$. This result is consistent
with Monte Carlo results of Selke et al. \cite{59}.

 For $D_t=1.965 < D \leq2$ the transition of the Blume-Capel model is believed to be first
 order, and the emerging interfaces $A|DO$ and $DO|B$ remain sharp as $T \rightarrow T_c(D)$.
 Reducing the problem to interfaces in the solid-on-solid model, one can argue that
 Eq.~(\ref{eq9}) also holds in the first-order region, but $\omega=1/3$ \cite{12}. These considersations
 have been checked by early Monte Carlo work \cite{58,59,60} for interfaces between coexisting
 bulk phases, but no study in the context of wetting at external boundaries has as yet been
 performed, to our knowledge.

In the present work, we wish to contribute filling this gap and study wetting behavior
 for the two-dimensional Blume-Capel model in the case of thin films for which surface 
 fields at the boundaries act, as described in Sec. 2. However, in order to do so, we have found it necessary 
 to reconsider the simulation methodology for the study of critical wetting. In fact,
 by applying Monte Carlo simulations, we shall necessarily study finite systems, namely strips of
width $L$ (in $y$-direction) and length $M$ (in $x$-direction, where a periodic boundary condition acts).
As a consequence, finite size effects matter, and hence in the next section we also shall give the background
on the proper finite size scaling analysis of   such simulation ``data'', for the case of critical  wetting
in $d=2$ dimensions, taking into account that one deals there with an anisotropic critical phenomenon of a
special character: if we consider the magnetization $m$ of the strip as the ``order parameter'' of the
transition, its critical exponent $\beta=0$. This fact does not seem to have found much attention in the
previous Monte Carlo studies of critical wetting in $d=2$ \cite{28,38}. This revised methodology for 
the study of critical wetting, which is outlined in Sec. 2, is another important result of our work.

Sec. 3 then presents our Monte Carlo results, which then are analyzed according 
to these concepts, and the wetting transitions are
discussed for a variety of choices of $D$, both cases $D < D_t$, $D=D_t$ and $D > D_t$ will be considered.
Sec. 4 then discusses our result on interfacial adsorption, and Sec. 5 summarizes our conclusions.

\section{THEORETICAL BACKGROUND AND DETAILS ON SIMULATION METHODS}

\subsection{The model}

We consider the Hamiltonian of the $3$-state Blume-Capel model \cite{49,50} where each lattice
site $i$ carries a spin $S_i$ that can take on the values $S_i= \pm1, 0,$ at a square lattice 
in a $L \times M$ geometry, where periodic boundary conditions act in the $x$-direction 
(where the lattice is $M$ rows long), while
free boundary conditions are used in the $y$-direction, where boundary fields $H_1$, $H_L$ act on the first
and the last row. Thus the Hamiltonian is

\begin{eqnarray} \label{eq10}
&&\mathcal{H}=-J \sum\limits_{\langle i,j \rangle} S_iS_j + D \sum\limits_i S^2_i  \nonumber\\
&& -H \sum\limits_i S_i - H_1 \sum\limits_{i \in \,{\rm row} 1} S_i  - H_L \sum\limits_{i \in \, {\rm row} L} S_i \quad .
\end{eqnarray}

Here $J$ is the exchange constant between spins at nearest neighbor sites, which we take homogeneous
throughout the system. Thus, we disregard the possibility to take the exchange $J_s$ in the boundary
rows different from the exchange $J$ in the interior of the system, that is frequently considered in studies
of wetting phenomena in the Ising model \cite{18,25,26,31,32,33,39}. The bulk field $H$ acts on
all lattice sites, while the ``surface fields'' \cite{45,46,47} only act on the spins in the
first (1) and last (L) row, where the free boundary conditions apply.

We also specialize on the particular antisymmetric situation $H_1=-H_L <0$ and consider the
thermodynamic limit $(L \rightarrow \infty$, $M \rightarrow \infty$) before we consider
the limit $H \rightarrow 0^+$. Then, the system undergoes two phase transitions: at the
temperature $T_{cb}(D)$ the phase transition occurs from the disordered ``paramagnetic'' 
phase to the ordered ``ferromagnetic'' phase, where we have used the
terminology of the Ising model. Taking the lattice spacing as our unit of length, 
the total number of spins is just the product of the linear 
dimensions of the system, $N=LM$. In fact, when
we define the magnetization $m$ per lattice site as

\begin{equation} \label{eq11}
m= \frac{1}{N} \sum\limits_{i=1}^N S_i \quad,
\end{equation}

\noindent its thermal expectation value $\langle m \rangle_T$ for temperatures $T<T_{cb} (D)$ will
be nonzero and positive in the considered limit. However, a second phase transition occurs at a lower
temperature $T_w(H_1)$ for small enough absolute values $|H_1|$ of the surface field: Note
that the surface field $H_1$ is oppositely oriented to the positive bulk field $H$, but
$H \rightarrow 0^+$ while $H_1$, stays finite. Thus, near $T_{cb} (D)$ the surface field
stabilizes a macroscopically thick layer of negative magnetization near the lower boundary,
where $H_1 <0$ acts, separated by an interface from the bulk, where the magnetization is
positive. At $T_w (H_1)$, a transition occurs where this interface gets localized
near the lower boundary: in the extreme case, the domain with negative magnetization disappears
completely. In the Ising model, which results from Eq.~(\ref{eq10}) as the limiting case
$D \rightarrow - \infty$, this wetting transition is second order throughout the regime
$0<|H_1|< J$.

\subsection{Critical wetting in $d = 2$ dimensions: A brief review}

For the semi-infinite system described above, the wetting transition is a singularity of the surface excess
free energy $f_s^{(1)}(T;H, H_1)$, defined from standard decomposition of the total free energy
into the bulk term and boundary terms, for $L \rightarrow \infty$, $M \rightarrow \infty$

\begin{eqnarray} \label{eq12}
&&F(T,H,H_1, H_L, L,M)/(LM) = f_b (T,H) + \frac{1}{L} f_s^{(1)} (T,H,H_1) \nonumber\\
&& \hspace{4,8cm}+ \frac{1}{L} f_s^{(L)} (T,H, H_L) \quad .
\end{eqnarray}

The singular part of this boundary free energy, also called ``wall tension'' 
or ``wall excess free energy'' \cite{1,2,9,10,11,12,13,14,18}, 
near $T_w(H_1)$ is expected to satisfy a scaling behavior
\cite{12,21,57,61,62} where $t=1-T/T_w (H_1) \rightarrow 0$,

\begin{equation} \label{eq13}
f_{s, {\rm sing}}^{(1)}/k_BT =|t|^{2-\alpha _s} \tilde{F}_s(H|t|^{-\Delta_s})
\quad .
\end{equation}

\noindent $\tilde{F}_s$ being a scaling function that we do not specify here. This is analogous to
standard scaling in the bulk near $T_{cb}$ \cite{56}, where $\tau=1-T/T_{cb} \rightarrow 0$,

\begin{equation} \label{eq14}
f_{b, {\rm sing}} / k_BT = |\tau|^{2- \alpha} \tilde{F}_b (H |\tau|^{- \Delta}) \quad .
\end{equation}

In Eq.~(\ref{eq14}), $\alpha$ is the specific heat exponent, and the ``gap exponent'' $\Delta$
is related to the standard exponents $\beta$ of the spontaneous magnetization $(\langle m \rangle_T \propto
\tau^\beta)$ and $\gamma$ of the susceptibility $(\chi=(\partial \langle m \rangle/\partial H)_T) \propto
|\tau|^{- \gamma}$ via the scaling relation $\Delta= \gamma + \beta$ \cite{56}. Thus $\alpha_s$,
$\Delta_s$ are analogous exponents characterizing the singular behavior of the critical wetting 
transition in $d=2$ dimensions.

As always when one considers a critical phenomenon, a diverging correlation length exists. Since wetting
can be viewed as an ``interface unbinding'' transition \cite{9,10,11,12,13,14,15,16,17,18} when
approached from below, it is natural to study the correlation function $C(x)$ describing the correlation of
fluctuations of the height of the contour $\ell(x)$ separating the domain with negative magnetization near
the lower boundary from the region with positive magnetization in the bulk. Defining $\delta \ell (x)$ as
$\delta \ell (x)= \ell(x)- \langle \ell \rangle$, this yields

\begin{equation} \label{eq15}
G(x) = \langle \delta \ell (0) \delta \ell (x) \rangle_T \quad,
\end{equation}

\noindent noting translational invariance in $x$-direction. 
By $\langle ... \rangle_T$ we denote an average in the canonical ensemble at temperature $T$;
in the following the subscript $T$ will be omitted. 
Near critical wetting in $d=2$ $G(x)$ takes the
scaling form \cite{61}

\begin{equation} \label{eq16}
G(x) = x^{-(1+\eta_{||})} g(x/\xi_{||}) \quad ,
\end{equation}

\noindent writing analogously to the decay of bulk correlations near criticality,
$G(r) = r^ {-(d-2+ \eta)} g(r/\xi_b)$, with $\eta$, $\eta_{||}$ being appropriate
exponents describing the variation of these correlations at large distances
right at bulk criticality or critical wetting, where $\xi_b$ (or $\xi_{||})$ is infinite.
For $\xi_{||}$, a scaling relation analogous to Eq.~(\ref{eq13}) holds,

\begin{equation} \label{eq17}
\xi_{||} = t^{-\nu_{||}} \tilde{\xi}_{||} (H t ^{-\Delta_s}) \quad ;
\end{equation}

\noindent again this Ansatz is inspired by the corresponding bulk behavior,

\begin{equation} \label{eq18}
\xi_b= \tau^{-\nu} \tilde{\xi}_b (H|\tau|^{- \Delta}) \quad,
\end{equation}

\noindent and again $\tilde{\xi}_{||}$, $\tilde{\xi}_b$ are scaling functions that we do not specify here.


We disregard here the problem that in general a nonlocal theory \cite{63,64,65,66}
is required \cite{40}, since it does not change the following conclusions. Capillary
wave theory implies that $\eta_{||}=0$, so there is no second independent critical exponent as in
the bulk. For a more detailed reasoning to explain why for the case of critical wetting with 
short-range forces the exponent $\eta_{||}$ is zero in all dimensions,
and hence there is a single independent critical exponent, we refer to the literature \cite{12}. 
Remember that using hyperscaling relations $d \nu=2 - \alpha=
\gamma + 2\beta$ and the scaling law $\gamma=\nu (2 -\eta)$ one can express
all static critical exponents just in terms of only two independent exponents
$\nu$ and $\eta$ in the bulk \cite{56}. For critical wetting, in $d=2$, there is
only one independent critical exponent namely, \cite{11}

\begin{equation} \label{eq19}
\nu_{||}=2
\end{equation}

\noindent and the hyperscaling relation for interfacial phenomena in $d=2$ dimensions $\nu_{||}=2
-\alpha_s$ \cite{12} then implies $\alpha_s=0$. From Eq.~(\ref{eq16}) we then conclude
that on a length scale $x$ the interface exhibits (at $t=0$) a mean-square displacement
also proportional to $x$, while for $t > 0$ one concludes that on the length scale
$\xi_{||}$ the mean-square displacement in $y$-direction also is proportional to $\xi_{||}$,
while for $x \gg \xi_{||}$ the correlation $G(x) \rightarrow 0$. This leads to the
conclusion that the correlation length $\xi_\bot$ describing the fluctuations of the interface
in $y$ direction scales as

\begin{eqnarray} \label{eq20}
&& \xi^2_\bot \propto \xi_{||} \quad , \quad \nonumber\\
&& \xi_\bot \propto t^{-\nu_\bot} ({\rm for}\, H=0), \nu_\bot=1 \quad .
\end{eqnarray}

Noting from Eq.~(\ref{eq13}) that surface excess magnetization $m_s$ and surface excess
susceptibility $\chi_s$ follow from derivatives with respect to the field

\begin{equation} \label{eq21}
m_s= - \partial f_s^{(1)} (T, H, H_1)/\partial H|_{T, H_1} \quad ,
\end{equation}

\begin{equation} \label{eq22}
\chi_s= - (\partial ^2 f_s^{(1)} (T, H, H_1)/\partial H^2)_{T, H_1}
\end{equation}

\noindent we conclude that these quantities have the singularities

\begin{equation} \label{eq23}
m_s \propto t^{2- \alpha_s-\Delta_s} \quad , \quad m_s \propto t^{\beta_s} \quad,
\end{equation}

\noindent and

\begin{equation} \label{eq24}
\chi_s \propto t^{2- \alpha_s - 2 \Delta_s} \quad \cdot \quad \chi_s \propto t^{- \gamma_s} \quad .
\end{equation}

Using then the consideration that this excess susceptibility is just caused by displacements
of the interface between the domain of negative magnetization at the boundary and the
bulk, one can use Eq.~(\ref{eq16}) to deduce a further scaling relation between
$\Delta_s,$ $\nu_{||}$ and $\eta_{||}$, namely \cite{12}

\begin{equation} \label{eq25}
\Delta_s=(\nu_{||}/2) [(d-1) + 2 - \eta_{||}] = 3, \quad (d=2) \quad.
\end{equation}

Also the excess magnetization $m_s$ obviously simply is related to $\langle \ell \rangle$,
and we have, in $d=2$,

\begin{equation} \label{eq26}
m_s \propto \langle \ell \rangle \propto t ^{- 1} \propto \xi_\bot \quad .
\end{equation}

Thus, near to critical wetting in $d=2$ the typical mean distance of the interface
from the wall and the typical excursions of this interface from its mean are of the
same order, $\nu_\bot=-\beta_s=1$.

\subsection{Finite-size scaling: wetting  vs bulk transitions}

All above relations did refer to the case that we were considering the limits
$M \rightarrow \infty$ and $L \rightarrow \infty$ first, and then consider
the limits $H \rightarrow 0$, $t \rightarrow 0$ near the wetting transition. While
such an approach is natural in the context of analytic theories \cite{10,11,12}, it it
not sufficient to understand simulations, where we wish to take these limits in reverse
order. When $L$ is kept large but finite, it is clear that for temperatures above
the wetting transition and $H=0$ there is no physical distinction between the domain
with negative magnetization in the lower half of the system and the domain with
positive magnetization in the upper half: Previously, for the semi-infinite case, the
positive magnetization in the bulk was singled out by taking the limit
$H \rightarrow 0^+$ first, and then the positive boundary field $H_L$ for
$L \rightarrow \infty$ was irrelevant, while for $L$ finite it is critical
to maintain the symmetry of the situation with respect to the sign of the
magnetization for $H=0$. As a consequence, we conclude that for the present situation
there is no bulk magnetization to consider, and it rather is the total magnetization
of the system or strictly speaking, its absolute value, which undergoes a transition
from zero to a nonzero value when we cross the wetting transition temperature, and
extrapolate simulation results towards $L \rightarrow \infty$.

As a consequence of this consideration, we propose a scaling assumption for the
distribution function $P_{L,M} (m)$ of the total magnetization in this finite
geometry as follows \cite{67,68}

\begin{equation} \label{eq27}
P_{L,M}(m) = \xi_{||}^{\beta/\nu_{||}} \tilde{P} (L^{\nu_{||}/\nu_\bot} /M, M / \xi_{||},
m \xi_{||}^{\beta/\nu_{||}}) \quad .
\end{equation}

Note that this expression generalizes the standard expression for finite-size scaling in isotropic
systems that have linear dimension $L$ in all spatial directions \cite{69},
as appropriate for phase transitions in the bulk

\begin{equation} \label{eq28}
P_L(m) = \xi_b^{\beta/\nu} \tilde{P}_b (L/\xi_b, m \xi^{\beta/\nu}_b)
\end{equation}

\noindent to systems with anisotropic linear dimensions $L,M$ and anisotropic correlation length
exponents $\nu_{||}$, $\nu_\bot$. While for isotropic critical phenomena
an anisotropic system shape would lead to a dependence simply on the ``aspect ratio''
$L/M$ of the system, the fact that $M$ scales with $\xi_{||}$ and $L$ scales with $\xi_\bot$
can be used to demonstrate \cite{67,68} that the $L$-dependence enters via ``the generalized
aspect ratio'' $L^{\nu_{||}/\nu_T}/M$ in the scaling function, Eq.~(\ref{eq27}).
With $\nu_{||}= 2$, $\nu_{\bot} = 1$ we simply have an argument $L^2/M$, which needs to be
kept constant when the variation with $M$ is studied. 
Such generalizations of finite-size scaling to anisotropic criticality have been discussed 
earlier for the Kasteleyn transition \cite{X,XI}, which also exhibits $\nu_{||} = 2 \nu_{\bot}$
and hence the same generalized aspect ratio $L^{2}/M$ applies.
The prefactor $\xi_{||}^{\beta/\nu_{||}}$ in
Eq.~(\ref{eq27}), and likewise the prefactor $\xi^{\beta/\nu}_b$ in Eq.~(\ref{eq28}), ensures
that the probability $P_{L,M}(m)$ can be properly normalized,

\begin{equation} \label{eq29}
\int\limits_{-1}^{+1} P_{L,M} (m) dm =1 \quad .
\end{equation}

Taking now suitable moments of $P_{L,M}(m)$, analogous to the procedure at isotropic phase
transitions in the bulk \cite{69,70}, we obtain

\begin{eqnarray} \label{eq30}
&& \langle |m| \rangle = \int\limits_{-1}^{+1} dm |m| P_{L,M} (m) = \nonumber\\
&& =\xi_{||}^{- \beta/\nu_{||}}  \tilde{m} (L^{\nu_{||}/\nu_\bot} /M, M/\xi_{||}) \quad ,
\end{eqnarray}

\begin{eqnarray} \label{eq31}
&&\langle m^{2k} \rangle = \xi_{||}^{-2 k \beta/\nu_{||}} \tilde{m}_{2k} (L^{\nu_{||}/\nu_\bot}/
M, M/\xi_{||}) , \quad \nonumber\\
&&k=1,2, \cdots \quad.
\end{eqnarray}

In particular, from Eqs.~(\ref{eq30}),~(\ref{eq31}) we derive the standard behavior for the
``susceptibility''. We denote it by $\chi'$ since $\langle |m|\rangle^2$ rather than $\langle m \rangle
^2$ is subtracted, see Ref. \cite{70} for a discussion. So, one has

\begin{equation} \label{eq32}
k_BT \chi'=LM (\langle m^2 \rangle - \langle |m| \rangle ^2) \quad,
\end{equation}

\noindent and hence

\begin{equation} \label{eq33}
k_BT \chi'=LM \xi_{||}^{- 2 \beta/\nu_{||}} \tilde{\chi} (L^{\nu_{||}/\nu_\bot}/M, M/\xi_{||}) \quad ,
\end{equation}

\noindent with the scaling function $\tilde{\chi}\equiv\tilde{m}_2 - (\tilde{m})^2$, omitting all arguments for
simplicity.

Now the key task is to identify the ``order parameter exponent'' $\beta$ in the context of this
description of a critical wetting transition. For this purpose, we note that the singular behavior
of $\chi'$ can only be due to the singular behavior of $\chi_s \propto t^{-4}$, Eq.~(\ref{eq24}), where
we used the results $\alpha_s=0$, $\Delta=3$. A key point to note is that $\chi'$ was normalitzed per spin,
relating to the total volume $LM$, while $\chi_s$ in Eq.~(\ref{eq24}) is taken relative to boundary sites, so
is only normalized by $M$. Hence we conclude that at $T_w$, we have, using $L^2/M=c$ in Eq.~(\ref{eq33}) to
eliminate $L$

\begin{equation} \label{eq34}
k_BT \chi' \propto  \sqrt{c} M^{3/2-2 \beta/\nu_{||}}
\end{equation}

\noindent irrespective of how the finite constant $c$ is chosen. Writing then

\begin{equation} \label{eq35}
k_BT \chi'|_{T_w}= k_BT \chi _s / L= k_BT \chi_s M^{-1/2} c^{-1/2}
\end{equation}

\noindent and using a finite-size scaling relation for $\chi_s$

\begin{equation} \label{eq36}
\chi_s=t^{-4} \tilde{\chi} (M/\xi_{||}) \propto \xi^2_{||} \tilde{\chi} (M/\xi_{||})
\propto M^2 \quad , T=T_w
\end{equation}

\noindent and hence we find

\begin{equation} \label{eq37}
k_BT \chi'|_{T_w} \propto M^{3/2}  (\propto L^{3} ) \quad .
\end{equation}

Comparison of Eqs.~(\ref{eq34}) and~(\ref{eq37}) yields a central result of this section,
namely

\begin{equation} \label{eq38}
\beta=0 \quad ,
\end{equation}

\noindent and Eq.~(\ref{eq37}) also can be interpreted in terms of the standard finite-size scaling
result

\begin{equation} \label{eq39}
k_BT \chi'|_{T_w} \propto M^{\gamma/\nu_{||}} \quad , \quad \gamma=3 \quad .
\end{equation}

Second order transitions with an exponent $\beta = 0$ are rather unusual; 
for another recent example see Jaubert et al. \cite{XII}.
In the present case, we can also understand it from the fact that
$\langle |m| \rangle$ must tend to $m_{b}$ in the thermodynamic limit 
in the partially wet phase, where the interface 
is bound to one of the walls up to the transition point, while in the wet phase 
for the considered limit $\langle |m| \rangle = 0$. It is interesting to note that 
these critical exponents also satisfy the usual scaling
relation with the gap exponent $\Delta_s$ introduced in Eq.~(\ref{eq13}),

\begin{equation} \label{eq40}
\gamma +\beta =\Delta_s=3 \quad ,
\end{equation}

\noindent as well as the hyperscaling relation for anisotropic bulk critical phenomena
in $d = 2$ dimensions, 

\begin{equation} \label{eq41}
\nu_{||} + \nu_\bot= 2 \beta + \gamma = 3 \quad.
\end{equation}

One should not confuse the exponent $\beta$ of the total magnetization, resulting in the 
limit $L \rightarrow \infty$, $M \rightarrow \infty$, $L^2 /M=c$, with the exponent $\beta_s$ of the surface excess
magnetization $m_s$ defined in Eq.~(\ref{eq23}): while $\langle m \rangle \rightarrow 0$
as $t \rightarrow 0$, $m_s$ diverges as $t \rightarrow 0$. Actually, we have the standard scaling
relations \cite{47}

\begin{equation} \label{eq42}
\beta_s= \beta - \nu_\bot \quad , \quad \gamma_s=\gamma + \nu_\bot \, , \, \alpha_s=\alpha + \nu_\bot
\end{equation}

\noindent with $\alpha=-1$, as is easily checked by noting
that $\gamma + 2 \beta= 2 - \alpha =3$.

A short manipulation of  Eq.~(\ref{eq33}) shows that it can be rewritten as 

\begin{equation} \label{eq33b}
k_BT \chi' = L^{\gamma/\nu_{\bot}} \tilde{\tilde{\chi}} (L^{\nu_{||}/\nu_\bot}/M, L/\xi_{\bot}) \quad ,
\end{equation}
\noindent where $\tilde{\tilde{\chi}}$ is another scaling function. For a fixed generalized
aspect ratio  $L ^{\nu_{||}/\nu_T}/M (= c)$ this function has a maximum at same value $X_{max}$ of the 
argument $X \equiv L/\xi_{\bot}$. This implies that the height of the maximum scales 
as $k_BT \chi_{max}^{'} = L^{\gamma/\nu_{\bot}} (= L^{3}$ in the present case); see also Eq.~(\ref{eq37});
and its position $T_{max}$ scales as 
\begin{equation} \label{eq33c}
T/T_{max} -1 \propto  L^{-1/\nu_{\bot}} (=L^{-1}) . 
\end{equation} 
\noindent Incidentally, this extrapolation has already tentatively been used in an 
early work \cite{27} without detailed justification.

\begin{figure}
\centerline{
\includegraphics[width=7cm,height=7cm,angle= 00]{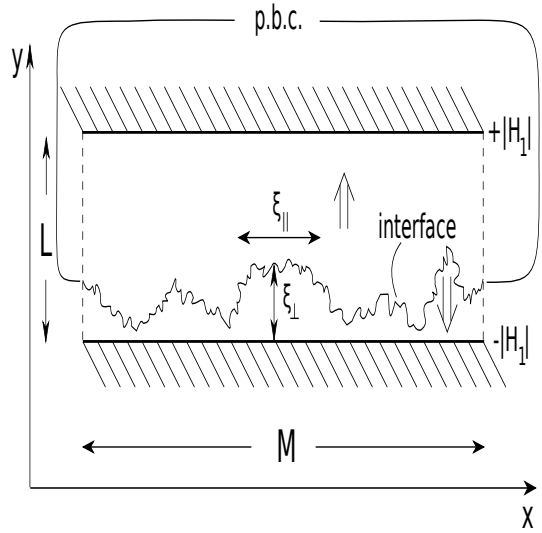}
}
\caption{\label{fig1} Schematic description of the system geometry and its state
slightly below the wetting transition temperature $T_w(H_1)$ such that both $\xi_{||}$
and $\xi_\bot$ are much larger than the lattice spacing. Coarse-graining the local
magnetization on a length scale intermediate between the lattice spacing and
$\xi_\bot$, one is left with one coarse-grained contour (the interface) separating
the domain with positive magnetization, which was assumed to be the majority domain
in the figure, without loss of generality, so the interface is still bound to the
lower wall, from the domain with negative magnetization. The sign of the magnetization
of the domains is indicated by double arrows. Note that $\xi_{||} \propto \xi^2_\bot$,
and the mean distance of the interface from the nearest boundary also is of the same
order as $\xi_\bot$. The choice of linear dimensions $L,M$, and 
of the periodic boundary conditions in $x$-direction
(p.b.c.) is indicated.}
 \end{figure}

The interpretation of the wetting transition in analogy to bulk critical phenomena,
considering a particular way of taking the thermodynamic limit (cf.~Fig.~\ref{fig1}), is
reminiscent of studies of filling transitions in double wedge \cite{71,72} or bipyramid
\cite{73,74} geometries; however, there the critical exponents are different. A useful
consequence of Eq.~(\ref{eq38}) is that both $\langle |m| \rangle$ and $\langle m^{2k} \rangle$ right at
the critical wetting transition become completely independent of linear dimensions, but
depend still on the constant $c=L^2/M$, the generalized aspect ratio. Plotting
$\langle |m| \rangle$ vs. temperature for different choices of $M$ at constant $c$,
critical wetting should show up via an unique intersection point. The fact that
for wetting transitions in $d=2$ for the choice $M/L^2 = {\rm const.}$ a unique intersection point
of $\langle |m|\rangle$ versus $T$ curves occurs has been only noted previously \cite{75} in a
study where long range boundary fields were applied, where $\nu_{||} = \infty$ holds \cite{75},
and hence the critical behavior is rather anomalous. Fig.~\ref{fig1} shows a schematic sketch
where the typical excursion $\xi_\bot$ of the interface is still smaller than $L$, by a factor of about 2.5,
and similarly $\xi_{||}$ is smaller than $M$. The absolute value $\langle |m| \rangle$ of the
magnetization then still is of the same order as the bulk magnetization $m_b$ that one encounters
inside of the domains. The distribution of $m$ then is distinctly bimodal, and $\langle m \rangle =0$,
since the interface is bound to the upper wall with the same probability as the lower wall. However,
when $T$ approaches $T_w$, the attractive force between the wall and the interface is compensated by
the entropic repulsion, the excursions of the interface extend over the whole width $L$, and then show
a correlation over a lateral length $M \propto L^2$. This interface configuration at $T_w(H_1)$ then is
selfsimilar, irrespective how large $L$ is chosen, as expected for critical phenomena. Thus, the
result $\beta=0$ must not be confused with a first-order wetting transition, where one would have
a discontinuous change from a bound to an unbound state of the interface, rather than this continuous
unbinding.

Of course, it is of interest to ask how the behavior changes when one does not take the thermodynamic
limit in this particular way where the generalized aspect ration $L^{2}/M$ is fixed, but rather choosing one of
the linear dimensions large but finite, and varying only the magnitude of the other linear dimension.
Thus, if $L$ is kept fixed and one considers the limit $M \rightarrow \infty$, the system becomes quasi-one-
dimensional. While mean-field theory would predict a sharp interface localization-delocalization transition
in this limit \cite{30}, one expects that thermal fluctuations cause a rounding of this transition
since the system is equivalent to a one-dimensional Ising system \cite{30}, and this expectation has in
fact been verified numerically \cite{76}. If instead we keep $M$ finite and increase $L$, the growth
of $\xi_{||}$ would be limited by the magnitude of $M$, and hence also $\xi_\bot$ as well as the
distance of the interface from one of the walls in the bound state would be limited by a value
proportional to $\sqrt{M}$, irrespective of $L$, while in the wet state its equilibrium position is at
$L/2$. As a consequence, data taken in such a way would appear as if the wetting transition were weakly of
first order, which it is not. Thus, we conclude that only the finite-size scaling analysis at
constant generalized aspect ratio $L^2/ M= {\rm const.}$ is the most useful choice.

\subsection{Comments on the simulation parameters}

In most of the numerical work, we have chosen a particular value of this 
generalized aspect ratio, namely

\begin{equation} \label{eq43}
L^2/ M = c= 9/8 \quad .
\end{equation}

Of course, the value of the constant $c$ in principle is arbitrary, and the results on the location
of $T_w(H_1)$ and the critical exponents should not depend on this choice. Our motivation for this
choice was simple that it yields solutions for $L$ and $M$ which are both integer, as it must be on the
lattice, and rather small: $(L,M)=(6,32); (12,128); (18, 288); (24,512); (30, 800)$; and $(36, 1152)$.
However, the effect of considering other choices of $c$ will be discussed below.

Monte Carlo simulations were then performed using the standard Metropolis algorithm, see e.g. \cite{70} for a
review. Typical runs are performed over a length of 10$^7$ Monte Carlo steps per lattice site (MCS),
disregarding the first $2 \times 10^6$ MCS to allow the system to reach equilibrium. Note that for systems
far below bulk criticality exposed to boundary fields cluster algorithms do not present any advantage \cite{44}.

\begin{figure}
\centerline{
\includegraphics[width=7cm,height=7cm,angle= 00]{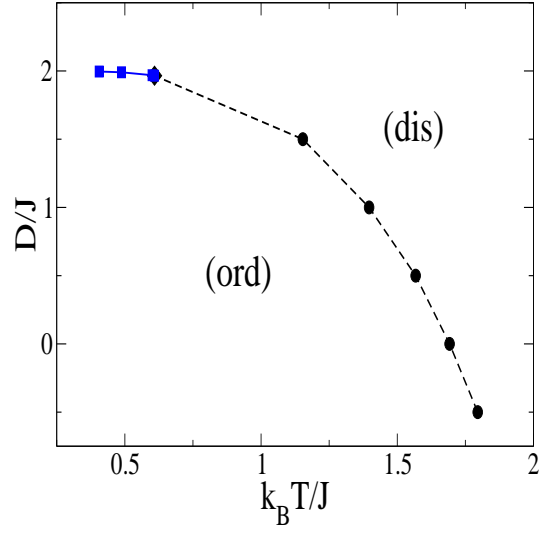}
}
\caption{\label{fig2} (Color online). Monte Carlo estimates for the locations of the phase boundary of the two-
dimensional Blume-Capel model, in the plane of variables $k_BT/J$ (abscissa) and $D/J$ (ordinate).
Full squares denote first-order transitions, the full diamond denotes the tricritical point
($k_BT_t/J \approx 0.609, \, D_t/J\approx 1.965)$, and full circles denote second-order transitions.
Broken straight lines connecting these points are guides to the eye only. The ordered ferromagnetic
phase ({\bf ord}) occurs below the line, and the disordered phase ({\bf dis}) above it,
as indicated. Note that the standard two-dimensional Ising model results in the limit where 
$D/J$ tends to minus infinity; so its transition is beyond the scale of the diagram.
}
 \end{figure}

In addition, part of the bulk phase diagram of the Blume-Capel model has been 
calculated, by applying again standard Monte Carlo methods \cite{70} but
using finite-size scaling for $L \times L$ lattices with periodic boundary conditions in both
$x$ and $y$ directions, see Fig.~\ref{fig2}. For earlier studies see e.g.~\cite{77,78}. 

\section{MONTE CARLO RESULTS ON THE WETTING BEHAVIOR}

\begin{figure}
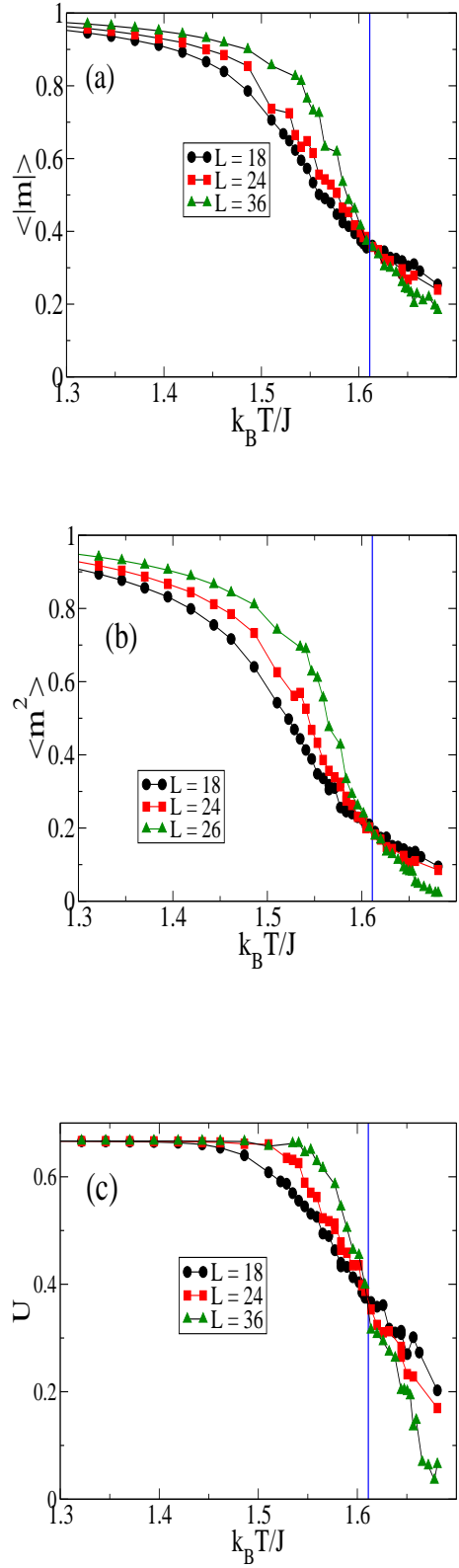

\centerline{
\includegraphics[width=6cm,height=6cm,angle= 00]{Fig3a.eps}
}
\vskip 1.0 true cm
\centerline{
\includegraphics[width=6cm,height=6cm,angle= 00]{Fig3b.eps}
}
\vskip 2.0 true cm
\centerline{
\includegraphics[width=6cm,height=6cm,angle= 00]{Fig3c.eps}
}
\caption{\label{fig3} (Color online). Plots of the average absolute value of the magnetization $\langle |m|\rangle$ (a),
the magnetization square (b), and  the cumulant (c) versus
temperature, for the choice $D/J=-\infty$, $H_1/J=0.70$, thus all data shown refer to the standard 
two-dimensional Ising model. The vertical lines indicate the exactly
known \cite{19} location of the wetting transition temperature.
Curves connecting points are drawn as guides to the eye (also in Figs.~\ref{fig4},~\ref{fig5}).}
 \end{figure}

\begin{figure}
\centerline{
\includegraphics[width=6cm,height=6cm,angle= 00]{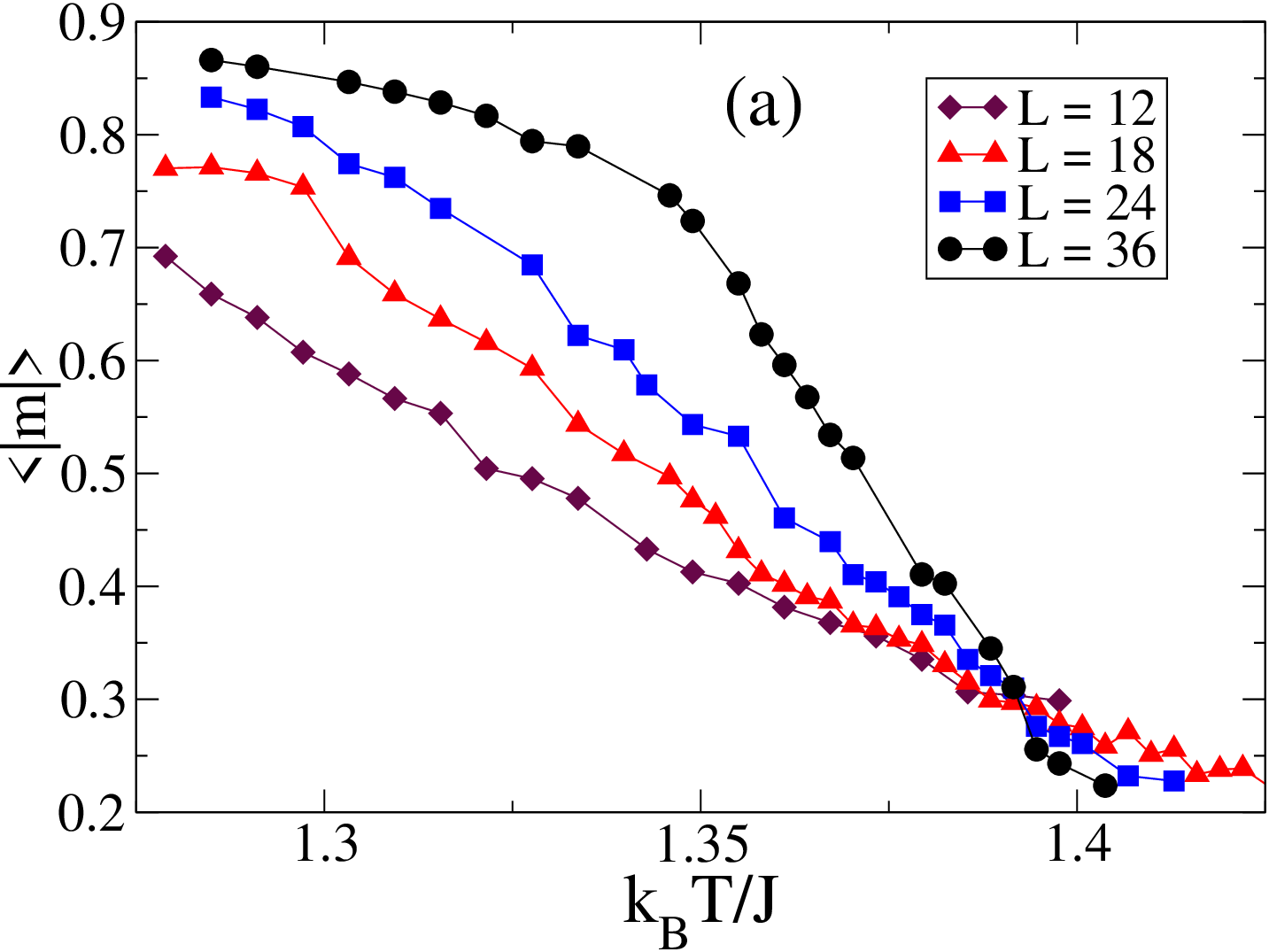}
}
\vskip 1.0 true cm
\centerline{
\includegraphics[width=6cm,height=6cm,angle= 00]{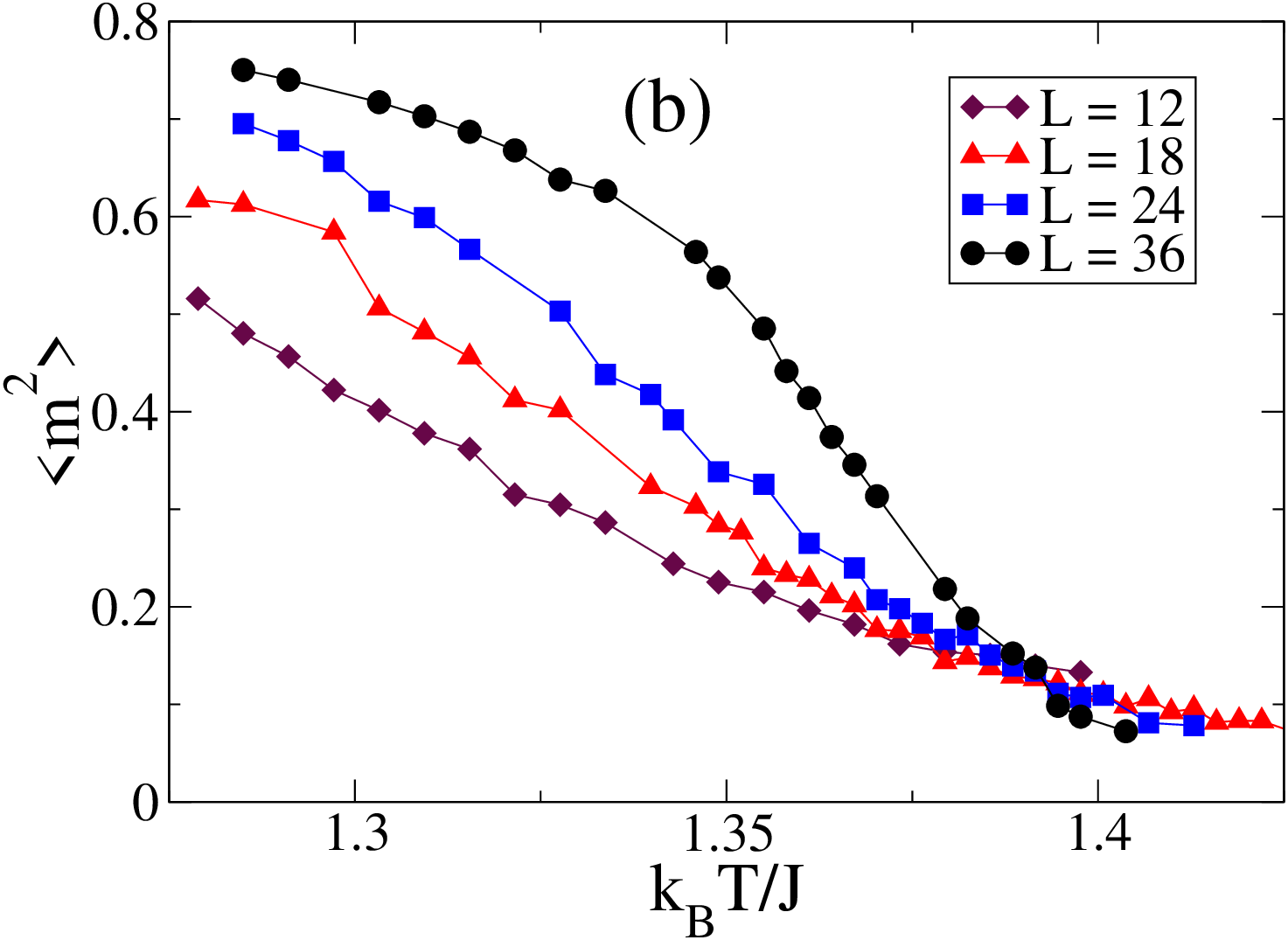}
}
\vskip 2.0 true cm
\centerline{
\includegraphics[width=6cm,height=6cm,angle= 00]{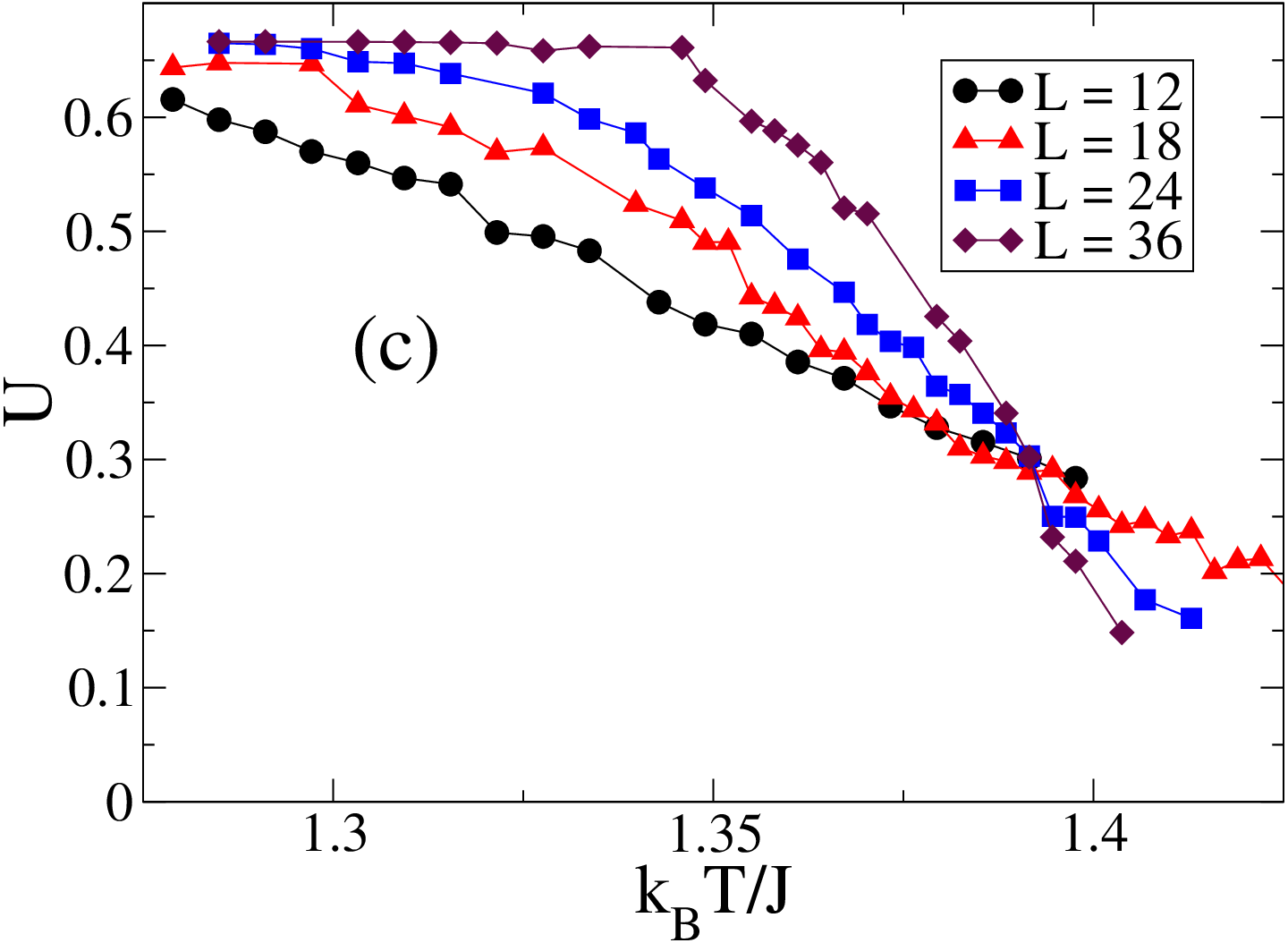}
}
\caption{\label{fig4} (Color online). Same as Fig.~\ref{eq3}, but for $D/J= 0.0$, 
$H_1/J=0.55$. From the intersection
points one can conclude $k_BT_w (H_1)/J = 1.393 \pm 0.004$, where the error bar
merely reflects the scattering of the intersection points of the measured observables.}
 \end{figure}

Motivated by the analysis of Sec. II, we have analyzed the first two moments
$\langle |m|\rangle$, $\langle m^2 \rangle$ and the cumulant \cite{69,70}

\begin{equation} \label{eq44}
U(T) = 1 - \langle m^4 \rangle /[3 \langle m ^2 \rangle^2 ]  \quad ,
\end{equation}

\noindent of the magnetization distribution and plot them versus temperature $T$ for a
few typical values of $D/J$, as shown in Figs.~\ref{fig3}-~\ref{fig5}). As expected,
rather well-defined intersection points in all three quantities
$\langle |m|\rangle$, $\langle m ^2 \rangle$ and $U(T)$ can be found at the
same estimate $k_BT_w(H_1)/J$ within
reasonably small errors. As always \cite{69,70}, the statistical accuracy
is better for $\langle |m|\rangle$ and $\langle m ^2 \rangle$ than for
the cumulant.
For standard phase transitions in the bulk \cite{70}, $\langle | m| \rangle$ and
$\langle m ^2 \rangle$ would not exhibit ``universal'', size-independent,
intersection points, of course. So, the behavior seen in
Figs.~\ref{fig3}-~\ref{fig5} is clear evidence for the scaling description 
of Sec. II, and also the direct observation
of configuration snapshots, see e.g.~Fig.~\ref{fig6}, is compatible with the
qualitative picture that was developed (Fig.~\ref{fig1}). Note that in the wet phase the average 
position of the interface is in the middle of the $L \times M$ strip,
in between the rows $y = L/2$ and $y = 1 + L/2$, for $L$ even as chosen here; 
but due to capillary waves on the scale $M$ in $x-$direction the interface makes excursions 
of order $\sqrt{M}$ in $y-$direction, which are of the order of $L/2$, for our choice 
of geometry (Eq. (\ref{eq43})).

On the other hand, in the case shown for $D = - \infty$, i.e. the Ising limit of
the Blume-Capel model (Fig.~\ref{fig3}), the location of the wetting transition
is exactly known \cite{18}, Eq.~(\ref{eq2}), and highlighted by a vertical straight line.
Evidently, the intersection of the curves in Fig.~\ref{fig3} occurs at a temperature compatible
with this prediction.

We do expect that the critical wetting transition of the Blume-Capel model falls in the same universality
class as for the Ising model. Thus, for the same choice of the generalized aspect ratio 
$c$, Eq.~(\ref{eq43}), we expect the same ordinate values of the intersection points of
$\langle |m|\rangle$, $\langle m ^2 \rangle$ and $U(T)$, respectively. Within our accuracy,
the data are compatible with this expectation.  

Fig.~\ref{fig7} analysis the scaling of the peak heights and peak locations of the susceptibility
$k_BT \chi'$, cf. Eq.~(\ref{eq37})  and Eq.~(\ref{eq33c}), respectively. 
Also, Figs.~\ref{fig8} and ~\ref{fig9} present some examples for the ``data collapse'' on 
master curves when $\langle |m| \rangle $, $\langle m^2 \rangle$ and $U$ are plotted as 
a function of $t \sqrt{M}$ as a test of Eq.~(\ref{eq30}),
$\langle |m| \rangle = \tilde{m} (L^2 /M$, $Mt^2 \hat{\xi}^{-1}_{||}$) where we used the asymptotic
power law for $\xi_{||}$, $\xi_{||} = \hat{\xi}_{||} t^{-\nu_{||}} =\hat{\xi}_{||}t^{-2}$. One sees
that a fair data collapse does in fact occur, although deviations due to both statistical errors
and systematic effects are present, when $M$ and/or $\xi_{||}$ are not large enough or when data too far away
from $T_w(H_1)$ are included. 
Of course, Figs.~\ref{fig7}-\ref{fig9} are just examples only, but representative for the general pattern 
of behavior. In any case, we
do conclude that using such finite-size scaling analyses as presented here one can locate wetting
transitions for two dimensional lattice models, such as the Blume-Capel model on the square lattice, with
reasonable accuracy.

On view of the well-defined intersection points and the data collapse in all three quantities
$\langle |m|\rangle$, $\langle m ^2 \rangle$ and $U(T)$ so far studied, figures \ref{fig3}-\ref{fig5}
and  \ref{fig8}-\ref{fig9}, respectively, it is worth to analyze the scaling behavior of the 
distribution function $P_{L,M} (m)$ of the total magnetization given by equation (\ref{eq27}).
In fact, since one has $\beta = 0$ the prefactor and the third scaling argument are constants. Also,
just at $k_BT_w(H_1)/J$ (i.e. $t = 0$) the second scaling argument vanishes due to the divergence of the
correlation length $\xi_{||}$, so that the distribution function depends on the 
generalized aspect ratio $c = L^{\nu_{||}/\nu_\bot} /M$, equation (\ref{eq43}), of the sample only. Figures 
\ref{fig10} (a) and (b) show plots of $P_{L,M} (m)$  versus $m$ as obtained by keeping $c = 9/8$ and for the 
cases $D/J = -\infty$ and $D/J=1.50$, respectively. It might be expected that the curves of
the probability distribution would be independent of $D$, except for normalization corrections due to the 
different density of vacancies. Since we found that the shape of the probability distribution depends
sensitively on $T$, it could be that the qualitative differences between the curves shown in figures
\ref{fig10} (a) and (b) would be due to the uncertainties in the location of $k_BT_w(H_1)/J$ for the case 
$D/J=1.50$.  

On the other hand, by keeping the temperature, the crystal field, and the surface magnetic field just 
at the wetting transition point, we have measured the dependence of $P_{L,M} (m)$  on the
generalized aspect ratio, for the Ising model, as shown in figure \ref{fig11} (a). In that figure we choose $ L = 30$ while $M$ is 
varied.  We recall the non-trivial shapes of the probability distributions that exhibit a single peak around 
$m = 0$ for rather elongated samples (i.e. $c \leq 0.75$), which monotonically crosses over to a 
bimodal distribution in the limit of more 'cubic' samples (i.e.  $c \geq 1.80$). Also, the generalized 
aspect ratio  
selected for our detailed Monte Carlo simulations ($c = 9/8$), which can roughly be located within the crossover 
regime, exhibits not only the central peak, but it also shows the onset of growth of lateral 
shoulders close to $m \simeq \pm 0.60$. By using the distributions already shown in 
figure \ref{fig11} (a)
we can also measure the dependence of the crossing points on the generalized aspect ratio of the sample 
(figure ~\ref{fig11} (b)). 
From this plot it follows that the suitable range of $c$ in order to obtain a reliable intersection point for both 
$\langle |m|\rangle$  and $U(T)$ in Monte Carlo simulations is rather narrow, say $0.75 \leq c \leq 2$, and 
slightly broader for the case of $\langle m ^2 \rangle$.
In order to be able to locate intersections with reasonable accuracy, it is advisable that the
cumulant intersection is neither close to zero nor close to its maximun value ($2/3$).   
It is also reassuring to see that for no shape the distribution at the wetting transition 
resembles the shape expected for a first-order wetting transition, which would be three delta 
functions at zero and positive and negative spontaneous magnetization in the thermodynamic limit,
and a distribution with three peaks of approximately Gaussian shape in a large but finite system.

The location of the wetting transitions as a function of two parameters, $H_1/J$ and $D/J$,
by determining well-defined intersection points 
still requires extensive computations and hence is a challenging task, Figs.~\ref{fig12} and ~\ref{fig13}.
Since $H_{1c} (J)$ at $k_BT_t/J=0.609$ decreases only rather slowly with increasing $D$ when one approaches
$D_t$, it obviously is very difficult to estimate the exponent $\Delta_{1t}$ (Eq.~(\ref{eq4}) numerically. 
In mean-field theory, supposed to be exact for $d \geq 4$ we 
know that $\Delta_1=1/2$ and $\Delta_{1t}=1/4$ \cite{47,52,79}; 
while in $d=3$ dimensions numerical estimates yielded \cite{26} $\Delta_1 \approx 0.45$, and we know that in $d=2$
$\Delta_1=1/2$ again. So, we see that the exponent $\Delta_1$ never deviates much from its mean-field value,
if at all. If we speculate the same observation to be true for $\Delta_{1t}$, we would expect that the data
in Fig.~\ref{fig13} for $T=T_t$ vary as $(1-D/D_t)^{1/4}$, which is not unreasonable.

We also see that in the region where the transition in the bulk is first order, the wetting transition lines
$H_{1c}(D)$ indeed end at nonzero values $H^*_{1c} (D)$ at the bulk transition line, and these values decrease
as one approaches the tricritical point, qualitatively compatible with Eq.~(\ref{eq6}). Thus, we can sketch
the global wetting behavior as shown schematically in Fig.~\ref{fig14}; however, it would be premature
to attempt to estimate the exponent $\varsigma$: our data are clearly too limited for this purpose, and we
have not tried to extend our study, since we feel the incomplete knowledge on the precise location of the
bulk tricritical point \cite{77,78} would hamper such a study.

As a final problem we consider the enrichment of vacancies at the interface between oppositely magnetized
domains in the Blume-Capel model. This problem has been considered earlier by Selke et al. \cite{58,59,60} for the case
of free unbound interfaces, while here we consider the extension where the interfaces are confined
between competing walls and may undergo a wetting (interface unbinding) transition. Fig.~\ref{fig15}
shows plots of the profiles of the vacancy density (a) and the magnetization (b) for a situation of incomplete
wetting. One observes a clear enrichment of the vacancies in the interfacial region. The more the interface
unbinds from the wall, the more the peak of the vacancy concentration moves away from the wall (Figs.~\ref{fig15} a,b).
Note that for the low temperature shown, the magnetization in the bulk tends to $\langle m \rangle =-1$, i.e.~
the vacancy concentration inside the bulk is very small.

Close to the wetting transition, the interface gets detached from the wall, and then the interface is
located in the middle of the strip, at $z=(L+1)/2$ on average. However, since the interface strongly fluctuates
around its average position a straightforward measurement of the vacancy concentration profile would
yield an almost horizontal flat curve across the strip. So we have defined a local coarse-graining of the
interface, taking segments of length $\Delta x=8$ in the $x$-direction, and determining the local center
of mass $z= \tilde{\ell}(x)$ of the interface in each segment, we compute the vacancy distribution in each
segment separately, relative to the center in each segment, and superpose the distributions from the
individual segments such that their centers coincide. In this way one obtains vacancy profiles with a
clear peak in the center of the strip in the wet phase (Fig.~\ref{fig15}c). Finally, when one reaches
the phase boundary of the bulk, the disordered phase takes over in the film, and the vacancy concentration becomes (almost)
unity in the system, except near the boundaries where the surface fields still stabilize layers of up-spins
and down-spins, respectively (Fig.~\ref{fig15}d). Note, of course, that the details of the curves in
Fig.~\ref{fig15}c,d do depend on this coarse-graining length $\Delta x$ distinctly.

This ambiguity that results depend
 on the coarse-graining is avoided when one simply computes
the average fraction of vacancies in the system (Fig.~\ref{fig16}). While for
$|H_1|$ chosen such that the system is in the incompletely wet state up to the
transition in the bulk one sees also a clear jump in the vacancy concentration from a small
value to almost unity when this transition in the bulk occurs, a much more rounded behavior
is found when wetting at the boundaries occurs. The difference between incompletely wet states
and completely wet states is even more pronounced when one studies the magnetization fluctuation:
for wet states, this fluctuation is very large for almost all values of $D$ until the transition in the
bulk is reached (Fig.~\ref{fig16}b). This behavior is corroborated by an examination of snapshot
pictures (Fig.~\ref{fig17}).

Finally, Fig.~\ref{fig18} shows an attempt to study the divergence of the interfacial adsorption, that we
expect when we approach the bulk transition in the wet phase in the region where a first-order
transition occurs. However, it is seen that the effective slope of these data changes slightly when the transition
is approached, and also depends on $|H_1|/J$. As a consequence, we must conclude that the asymptotic
region where a universal exponent could be estimated has not been reached. In view of the strong and slow
fluctuations of the interface it is not straightforward to cope with this problem.

\section{Conclusions}

In this paper, we have presented a Monte Carlo study of the wetting behavior of the two-dimensional Blume-Capel
model, mapping out the surface critical field $H_{1w}(T,D)$ where critical wetting occurs for a broad range 
of parameters $T$ and $D$, both in the region where the transition of the Blume-Capel model in the 
bulk is of second order (cf. Fig. \ref{fig12})  and where is of first order (cd. Fig. \ref{fig13}). 
We suggest that $H_{1c}(T,D)$ behaves as $H_{1c}(T,D) \propto (T_{cb}(D) - T)^{1/2}$ for all $D < D_{t}$ and 
$H_{1c}(T \rightarrow T_{cb}(D)) = H_{1c}^{*}(D)$ for $D > D_{t}$, and propose a qualitative 
phase diagram in the space of the three variables surface field ($H_{1}$), temperature ($T$)
and ``crystal field'' ($D$), as sketched in figure \ref{fig14}. Thus, we present evidence for
the universality of the surface critical exponent $\Delta_{1}$ along the critical line
of the Blume-Capel model. However, the precise crossover 
behavior in the inmediate vecinity of the tricritical point $(D_{t}, T_{cb}(D_{t}))$ could not yet 
been studied: first of all, the location of this point needs to be more accurately determined,
and also more powerful numerical methods (rather than the straightforward Monte Carlo
simulation method just using the Metropolis algorithm) would be required.

Already for the task that we did achieve we encountered the necessity to reconsider the 
finite-size scaling approach to the study of wetting transitions, Sec. II.  We propose 
that by using a procedure where the linear dimensions  $L, M$ of the $L \times M$ strip are
varied such that the generalized aspect ratio $c = L^{2}/M$ is kept constant,
cf. Eq. (\ref{eq43}), it is possible to analyze the data in full analogy to the study of a bulk 
phase transition, where one uses antisymmetric surface fields ($H_{1} = - H_{L}$). Then,
simply the total magnetization $m$ in the system acts like as order parameter,
but the appropiate critical exponent is $\beta = 0$. Using the exactly known results for the 
Ising model in the square lattice, which is the limit of the Blume-Capel model 
for $D \rightarrow \infty$, as a test case, our new finite-size scaling
approach is nicely verified. Note that $H_{1c}(T, D \rightarrow \infty)$ and the critical 
exponents $\nu_{||} = 2$, $\nu_\bot= 1$,  $\Delta_{s} = 3$, $\alpha_{s} = 0$, $\beta_{s} = -1$, and 
$\gamma_{s} = 4$ for  critical wetting are known. Then, we propose the corresponding exponents when 
one treats the transition not as a surface free energy singularity of the semi-infinite 
system in the limit where the bulk field $H$ tends to zero, but as a bulk singularity
of the $L \times M$ system for $H \rightarrow 0$ in the limit $L  \rightarrow \infty$,
$L^{2}/M = c =$ constant. These expònents are  $\alpha = -1$, $\beta = 0$ and 
$\gamma = 3$.  We have shown that these exponents satisfy all 
expected scaling relations and we have verified them from our simulations. 
Use of this formulation of finite-size scaling for critical wetting in $d = 2$ dimensions
is a convenient and useful tool; it would be interesting to study the extension of this method 
to $d = 3$ dimensions where $\nu_\bot= 0$ (logarithmic growth) and this extreme shape 
anisotropy that our method then requires makes the task obviously very much harder.

Since the  Blume-Capel model differs from the Ising model by the presence of vacancies as a
third component, it is interesting to ask to what extent vacancies get enriched at the interfaces 
in this situation of interfaces confined by walls. We found that the situation is fully      
analogous to interfaces between coexisting phases in the bulk: in the second-order region  
the interfacial adsorption is finite, while in the first-order region the predicted divergence
of interfacial adsorption is found.

Of course, there are many cases where by changing a parameter the order of a phase transition
in the bulk changes from second to first order, and one can ask how wetting phenomena are affected.
We hope that the present work will stimulate further theroretical and experimental studies along such lines. 

{\bf Acknowledgement}: One of as (EVA) received support from the Alexander von Humboldt Foundation 
and from the  Schwerpunkts  f\"ur Rechnergestützte Forschungsmethoden in den 
Naturwissenschaften (SRFN), Germany. Also, the support of the 
the CONICET, ANPCyT and UNLP (Argentina) is greatly acnowledged.

\clearpage

\clearpage

\begin{figure}
\centerline{
\includegraphics[width=6cm,height=6cm,angle= 00]{Fig5a.eps}
}
\vskip 1.0 true cm
\centerline{
\includegraphics[width=6cm,height=6cm,angle= 00]{Fig5b.eps}
}
\vskip 2.0 true cm
\centerline{
\includegraphics[width=6cm,height=6cm,angle= 00]{Fig5c.eps}
}
\caption{\label{fig5} (Color online). Same as Fig.~\ref{fig3}, but for $D/J=1.75$, $H_1/J=0.85$. 
From the intersection
points one can conclude $k_BT_w (H_1)/J=0.538 \pm 0.004$, where the error bar
merely reflects the scattering of the intersection points of the measured observables.}
 \end{figure}

\clearpage

\begin{figure}
\centerline{
\includegraphics[width=7cm,height=7cm,angle= 00]{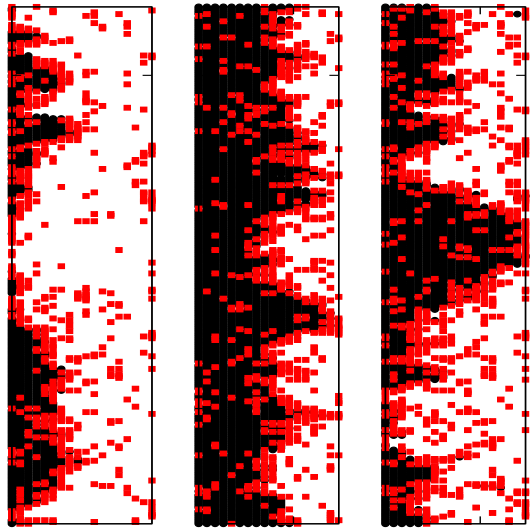}
}
\caption{\label{fig6} (Color online). Snapshot pictures of the spin configuration of the 
Blume-Capel model at $D/J=1.5$ and $H_1/J=0.7$ for $(L,M)=(18, 288)$ at three 
temperatures: $k_BT/J=0.445$ (left), 0.481 (middle) and 0.518
(right). The sites where $S_i=-1$ are shown in black, the sites where 
$S_i=0$ are shown in grey (red color online), while
sites with $S_i=+1$ are left blank. Note that the case $k_BT/J=0.481$ is close to the wetting transition.}
 \end{figure}

\begin{figure}
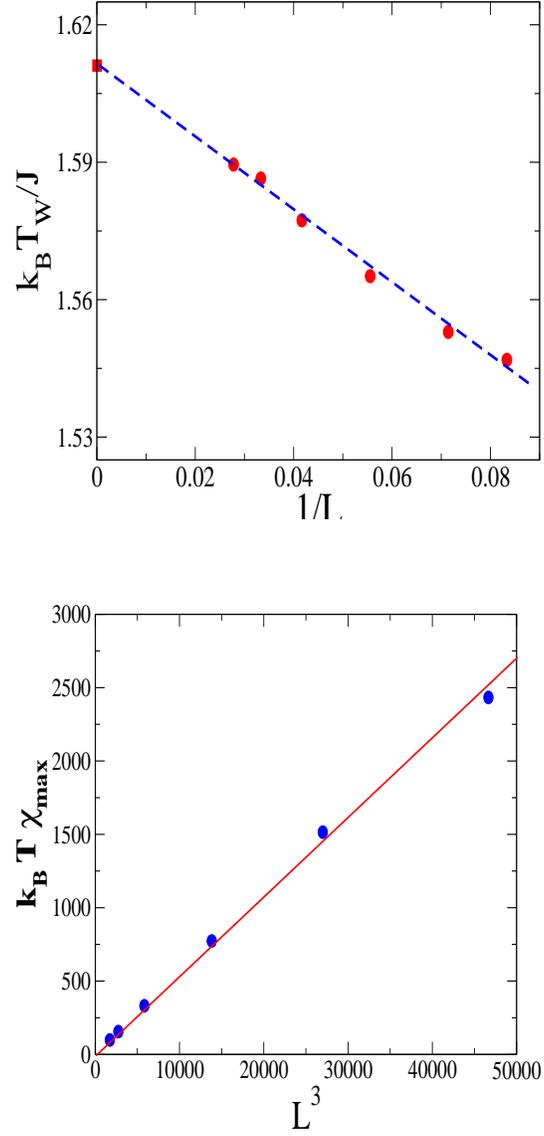

\centerline{
\includegraphics[width=7cm,height=7cm,angle= 00]{Fig7a.eps}
}
\vskip 1.0 true cm
\centerline{
\includegraphics[width=7cm,height=7cm,angle= 00]{Fig7b.eps}
}
\caption{\label{fig7} (Color online). a) Extrapolation of the 
peak position $k_BT_{\rm max}/J$ versus $1/L$ (see Eq.~(\ref{eq33c})) for the case
$D/J=-\infty, H_1/J=0.7$. The full square indicates the exactly
known \cite{19} location of the wetting transition temperature, and the 
dashed line has been drawn in order to guide the eye.
Note that $L^2/M=9/8$ (Eq.~(\ref{eq43})) is chosen throughout.
b) Linear-linear plot of $k_BT \chi_{\rm max}$ versus $L^{3}$, 
for the same choice as used in a), 
to demonstrate the power law $k_BT \chi_{\rm max} \propto
 L^{3} ~$ or $(M^{3/2})$, see Eq.~\ref{eq37}.}
\end{figure}

\begin{figure}
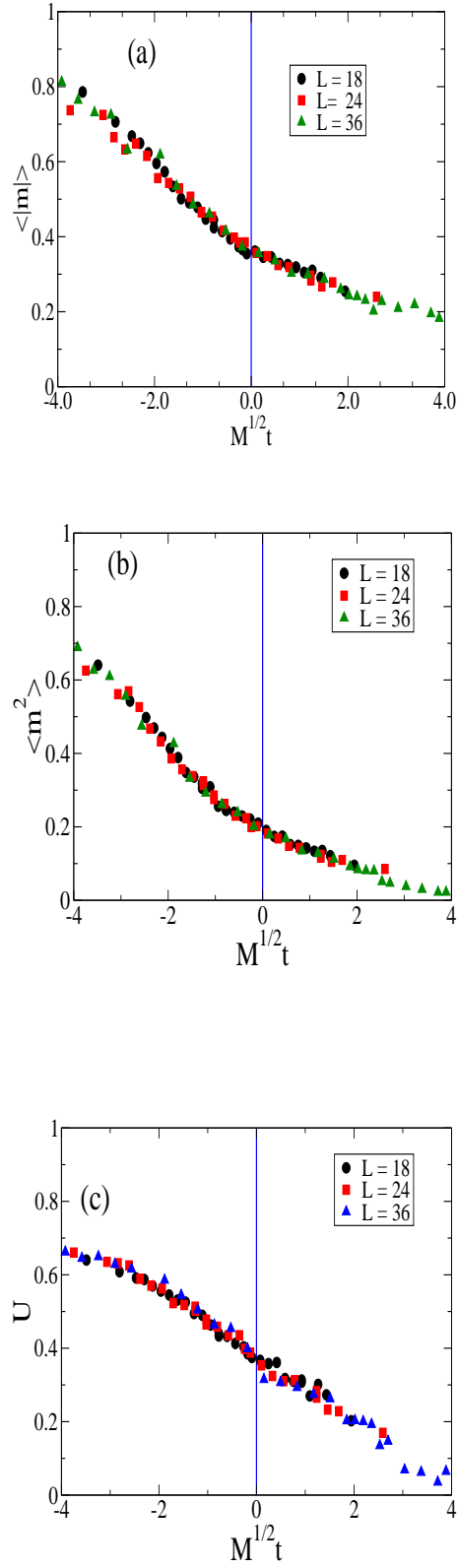

\centerline{
\includegraphics[width=6cm,height=6cm,angle= 00]{Fig8a.eps}
}
\vskip 1.0 true cm
\centerline{
\includegraphics[width=6cm,height=6cm,angle= 00]{Fig8b.eps}
}
\vskip 2.0 true cm
\centerline{
\includegraphics[width=6cm,height=6cm,angle= 00]{Fig8c.eps}
}
\caption{\label{fig8} (Color online). Scaling plot of $\langle |m| \rangle_T$ (a), 
$\langle m^2 \rangle $ (b), and $U$ (c)
versus $t\sqrt{M}$ for the same case $D/J=-\infty$,
$H_1/J=0.7$, and various choices of $L$ (for $L^2/M=9/8)$ as indicated in the figure.
The vertical line is a reminder that in this case the transition point 
$t = 0$ is known exactly.}
\end{figure}

\begin{figure}
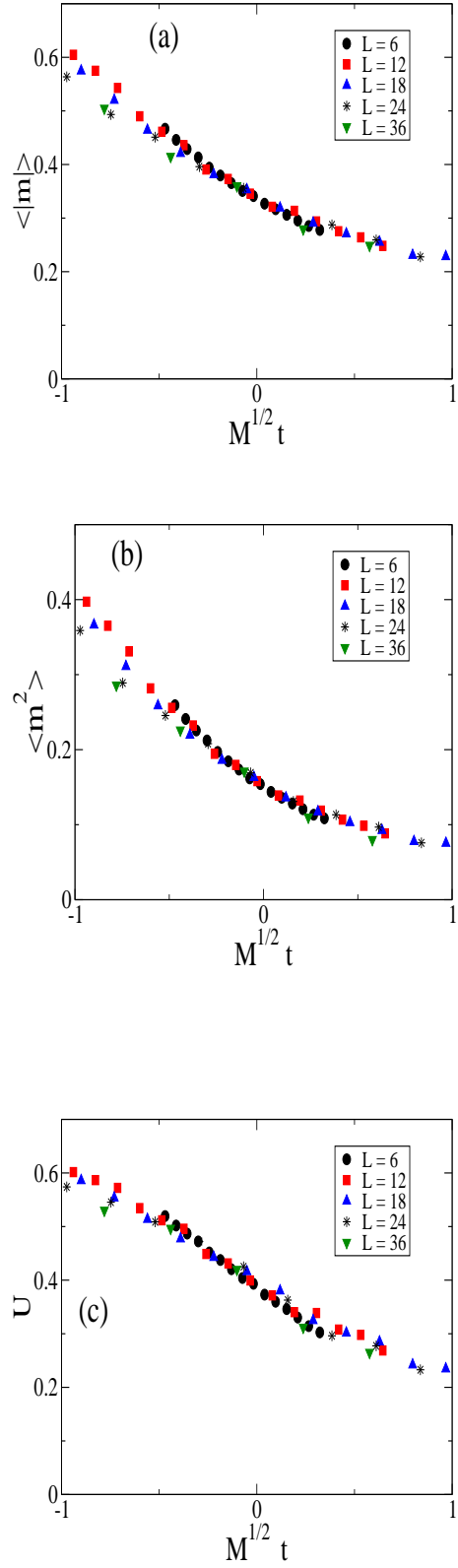

\centerline{
\includegraphics[width=6cm,height=6cm,angle= 00]{Fig9a.eps}
}
\vskip 1.0 true cm
\centerline{
\includegraphics[width=6cm,height=6cm,angle= 00]{Fig9b.eps}
}
\vskip 2.0 true cm
\centerline{
\includegraphics[width=6cm,height=6cm,angle= 00]{Fig9c.eps}
}
\caption{\label{fig9} (Color on line). Same as Fig.~\ref{fig8}, but for the case $D/J=1.75$, $H_1/J=0.85$.
The transition point $t = 0$ was determined in Fig.~\ref{fig5}.}
\end{figure}

\begin{figure}
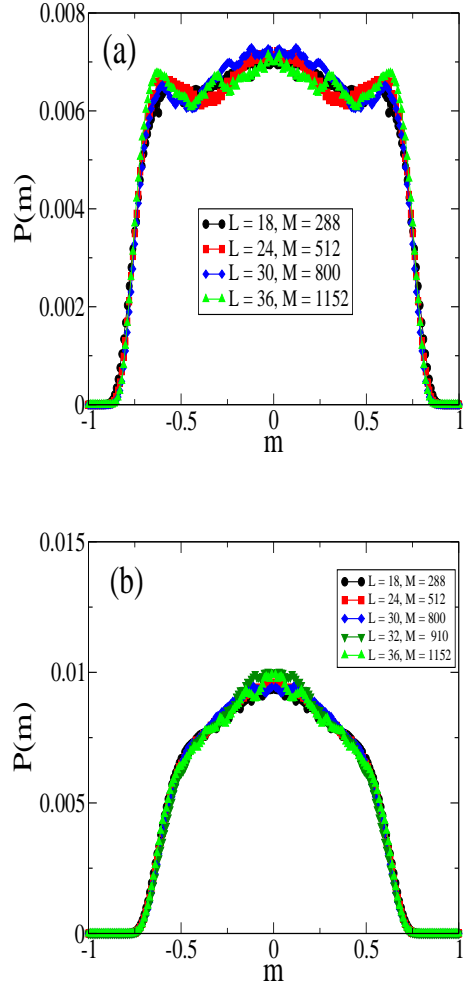

\centerline{
\includegraphics[width=6cm,height=6cm,angle= 00]{Fig10a.eps}
}
\vskip 1.0 true cm
\centerline{
\includegraphics[width=6cm,height=6cm,angle= 00]{Fig10b.eps}
}
\caption{\label{fig10} (Color online).
(a) Plots of $P_{L,M} (m)$  versus $m$ as obtained for  various choices of $L$
and $M$, as indicated, but keeping $L^2/M=9/8$ (Eq.~(\ref{eq43}))and for the case $D = -\infty$.
Data obtained for $H_1/J=0.70$ and the exactly known \cite{19} location of the wetting transition temperature,
namely $k_BT_w (H_1)/J \simeq 1.6111$. 
(b) As in (a) but for the case $D/J = 1.50$, and $H_1/J=0.70$. The best data collapse of 
the data is obtained for $k_BT_w (H_1)/J = 0.792$, in agreement with the intersection points
of the  quantities $\langle |m|\rangle_T$, $\langle m ^2 \rangle_T$ and $U(T)$,
which are not shown here for the sake of space.   
}
\end{figure}

\begin{figure}
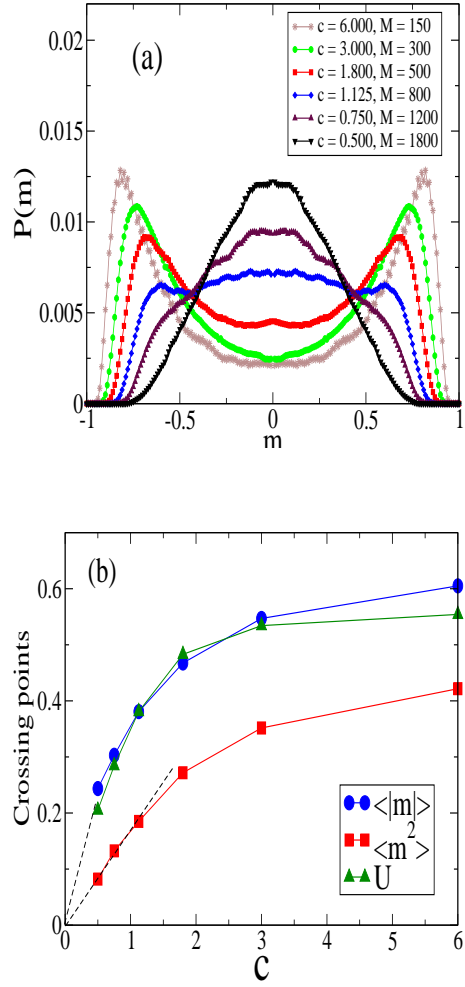

\centerline{
\includegraphics[width=6cm,height=6cm,angle= 00]{Fig11a.eps}
}
\vskip 1.0 true cm
\centerline{
\includegraphics[width=6cm,height=6cm,angle= 00]{Fig11b.eps}
}
\caption{\label{fig11} (Color online).
(a) Plots of $P_{L,M} (m)$  versus $m$ as obtained for $L = 30$ and various choices of $M$, 
in order to scan a wide range of the generalized aspect ratio $c = L^2/M$, as indcated.
Data obtained for $D/J = -\infty$, $H_1/J=0.70$ and the exactly known \cite{19} location of the 
wetting transition temperature, namely $k_BT_w (H_1)/J \simeq 1.6111$. 
(b) Intersection points of the  quantities $\langle |m|\rangle$, $\langle m ^2 \rangle$, and $U(T)$,
versus the sample generalised aspect ratio $c = L^2/M$ as obtained for the same parameters as in (a).
The dashed lines showing a linear extrapolation to $c = 0$ have been drawn  for the
sake of comparison. 
}
\end{figure}

\begin{figure}
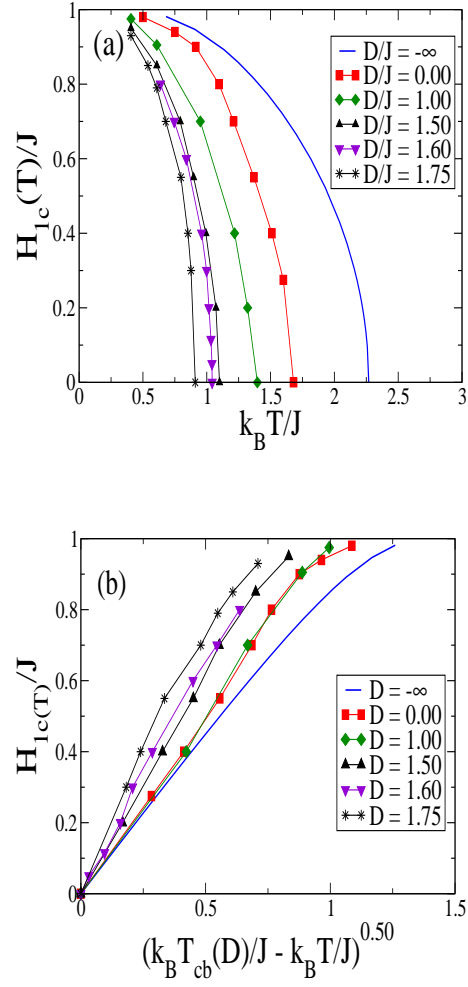

\centerline{
\includegraphics[width=6cm,height=6cm,angle= 00]{Fig12a.eps}
}
\vskip 1.0 true cm
\centerline{
\includegraphics[width=6cm,height=6cm,angle= 00]{Fig12b.eps}
}
\caption{\label{fig12} (Color online). a) Plot of $H_{1c}(T)/J$, the inverse 
function of $k_{B}T_w(H_1)/J$ versus temperature for a
range of values of $D/J$, as indicated in the figure. b) Plot of $H_{1c}(T)/J$ 
versus $(k_{B}T_{cb}(D)/J - k_{B}T/J)^{1/2}$
for a range of value $D/J$, to show that the exponent $\Delta_1 =1/2$, irrespective of $D/J$, in
the region of the second-order transition. In both cases a) and b), the full line corresponds to 
the exact solution for the case $D/J = - \infty$ given by Eq.~(\ref{eq2}) \cite{19}.}
\end{figure}

\begin{figure}
\centerline{
\includegraphics[width=7cm,height=7cm,angle= 00]{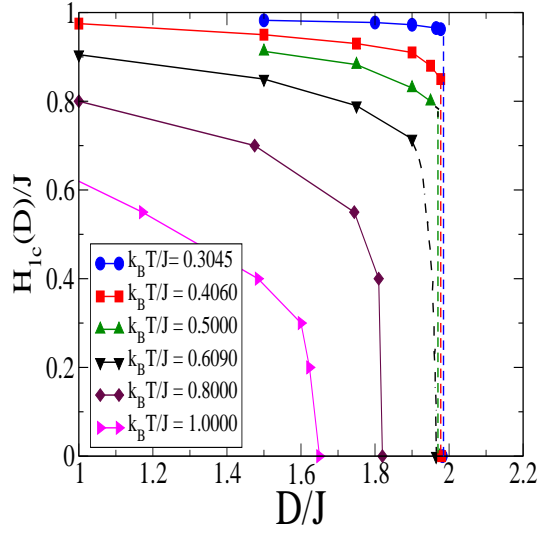}
}
\caption{\label{fig13}(Color online). Plot of $H_{1c}(T)/J$ 
versus $D/J$ for the tricritical temperature $(k_BT/J = 0.609)$, and
three temperatures in the region of $k_{B}T/J$ where the transition in the bulk clearly is first order, 
$k_BT/J=0.3045$,  $k_BT/J=0.406$, and  $k_BT/J=0.500$, respectively, as indicated. 
Also, two sets of data corresponding to the range of $k_{B}T/J$ where the transition 
in the bulk clearly is second order, $k_BT/J=0.8000$, and $k_BT/J=1.00$, respectively, 
are shown as indicated. Note that dashed-straight 
lines connecting the points are drawn as guide to the eye only.}
\end{figure}
\begin{figure}
\centerline{
\includegraphics[width=7cm,height=7cm,angle= 00]{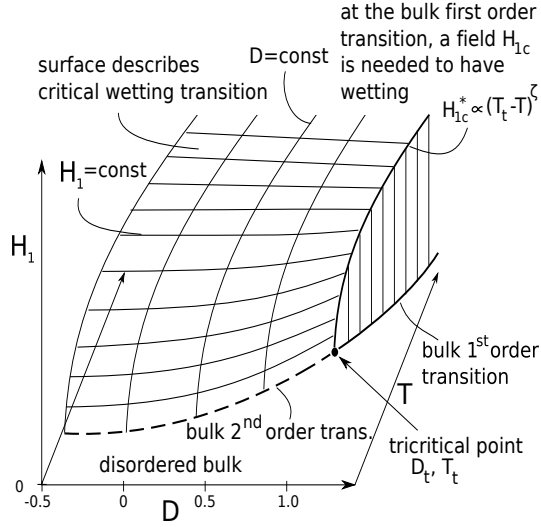}
}
\caption{\label{fig14}Schematic description of the wetting behavior of the two-dimensional Blume-Capel model in the space of 
variables $H_1$, $D$ and $T$. Complete wetting occurs above the surface of critical wetting transition, which ends 
at $H_1=0$ when the bulk transition is continuous, and at the line $H^*_{1c}(T)$ in the region beyond the tricritical 
point $(D_{t}, T_t)$, where the transition in the bulk is of 1$^{\rm st}$ order.}\end{figure}

\begin{figure}
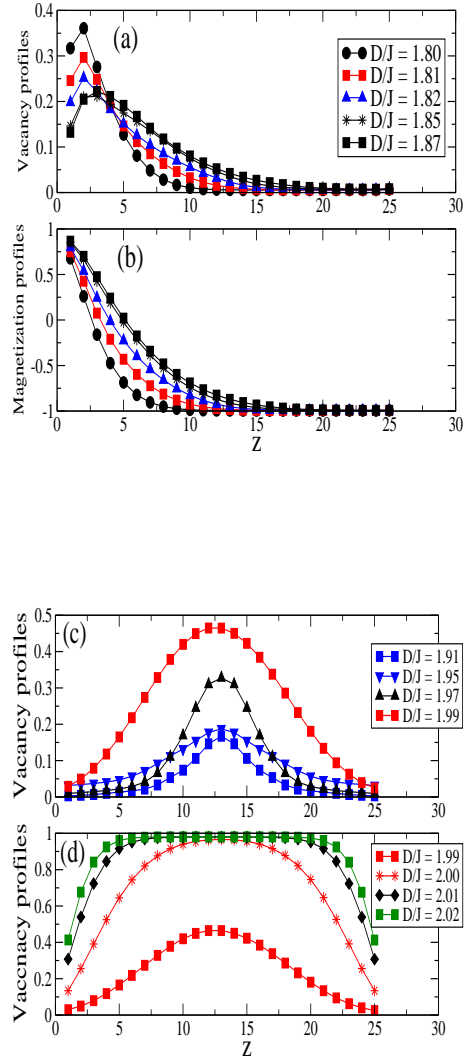

\centerline{
\includegraphics[width=6cm,height=6cm,angle= 00]{Fig15ab.eps}
}
\vskip 2.0 true cm
\centerline{
\includegraphics[width=6cm,height=6cm,angle= 00]{Fig15cd.eps}
}
\caption{\label{fig15}(Color online). Vacancy concentration profiles (a,c,d) and the magnetization profiles (b) plotted
as a function of the distance from the wall on which the positive surface field acts, in the case where the 
majority of the thin strip is in a state of negative magnetization in case (a), (b). Data refer to the
lattice with linear dimensions $(L,M)=(24, 288)$ and the reduced temperature $k_BT/J=0.406$, surface 
field $|H_1|/J=0.91$, and several choices of the parameter $D/J$, as indicated.}\end{figure}

\begin{figure}
\centerline{
\includegraphics[width=7cm,height=7cm,angle= 00]{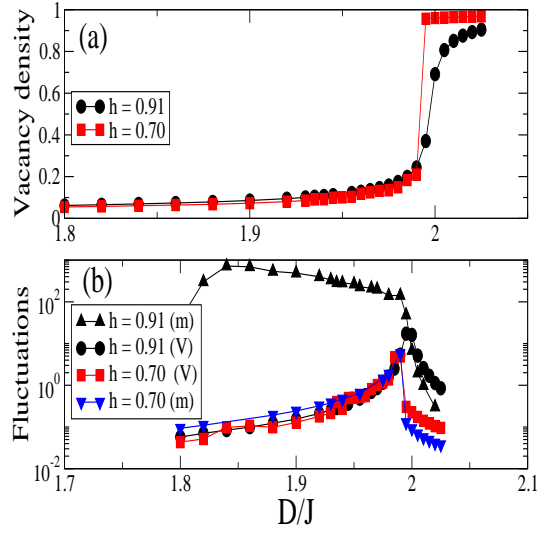}
}
\caption{\label{fig16}(Color online). Total concentration of vacancies in the system (a) and its fluctuation (b)
plotted vs. $D/J$ for $k_BT/J=0.406$ and several choices of the surface field $h = |H_1|/J$, as indicated.
Part(b) includes also the fluctuation of the magnetization $m$ for comparison. The system linear
dimensions were again chosen as $(L,M)=(24,288)$.}\end{figure}

\begin{figure}
\centerline{
\includegraphics[width=7cm,height=7cm,angle= 00]{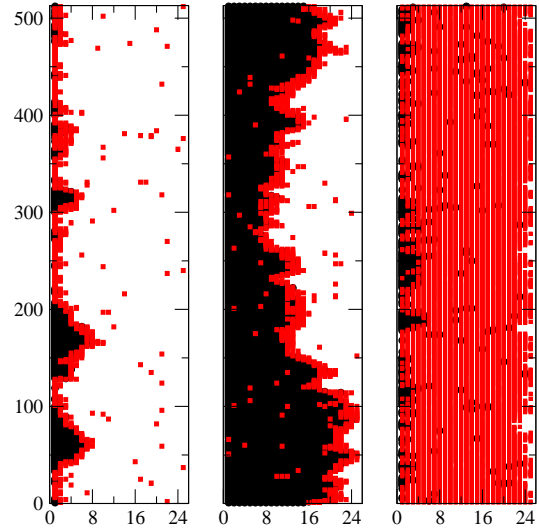}
}
\caption{\label{fig17}(Color online). Typical snapshot pictures of the spin configurations obtained for $k_BT/J=
0.406$, $|H_1|/J=0.91$ and three values of $D/J:D/J=1.80$ (left panel, refers to an
uncompletely wet state), 1.90 (medium panel, wet phase), 2.02 (right panel, disordered phase
beyond the first order transition of the bulk). Sites $i$ with $S_i=  1$ are shown by black
dots, sites $i$ with $S_i= 0$ are shown as grey (b/w) or red (in color) dots, 
sites $i$ with $S_i=-1$ are left blank.}
\end{figure}

\begin{figure}
\centerline{
\includegraphics[width=7cm,height=7cm,angle= 00]{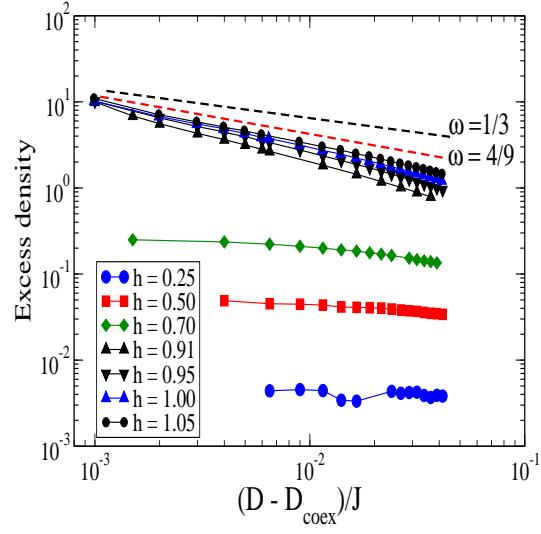}
}
\caption{\label{fig18} (Color online). Log-log plot of the excess density of the vacancies versus $(D_{\rm coex} -D)/J$
at $k_BT/J=0.406$ (where $D_{\rm coex}/J=1.996)$ and different values of the surface field $h = H_1/J$. Broken lines
show the theoretical slopes 1/3 (expected in the first-order region far off from the tricritical point)
and 4/9 (expected at the tricritical point), see also Eq.~(\ref{eq9}). Chosen lattice size was $(L,M)=(201,400)$.}
 \end{figure}

 \end{document}